Vladimir K. Ivanov,
Boris V. Palyukh
Tver State Technical University,
Afanasy Nikitin st. 22, Tver,
Russia
Email: {mtivk,
pboris}@tstu.tver.ru}

Alexander N. Sotnikov
Joint Supercomputer Centre of
Russian Academy of Science
Leninsky Prospect 32a, Moscow,
Russia
Email: asotnikov@jscc.ru

# Approaches to the Intelligent Subject Search

*Abstract*—This article presents main results of the pilot study of approaches to the subject information search based on automated semantic processing of mass scientific and technical data. The authors focus on technology of building and qualification of search queries with the following filtering and ranking of search data. Software architecture, specific features of subject search and research results application are considered.

I. INTRODUCTION

NEW efficient scientific knowledge search and synthesis methods (in particular, breakthrough technologies and innovative ideas in economics, science, education) are one of the top research and development targets in the field of information technology. The project Intelligent Distributed Information Management System for Innovations in Science and Education powered by the Russian Foundation of Basic Research is to solve this problem. This article presents main results of the pilot study of approaches to the subject information search based on automated semantic processing of mass scientific and technical data.

II. SPECIFIC FEATURES OF SUBJECT SEARCH

The major features of subject search tasks which determine the approaches are:

• the required information is often located at the junction of adjacent areas, hence, there is some complexity in the exact wording of the search query.

• along with the information on proper innovation it is desirable to obtain information on applications, risks, specific features, users, authors, producers.

• there is a necessity of available alternatives and different criteria mixing for selecting the most effective practices.

• the information on innovations is fragmentized and heterogeneous; primarily sector-specific character.

In contrast to the search for specific information (facts) on particular aspects of the required content, it is rather difficult to solve a sophisticated problem of searching coordinated information on a target subject. For example, it is required to find the economic performance of mine Raspadskaya JSCo for the first half of 2013. If we use this phrase as a search query, it is possible to get a relevant answer in the first ten search results of Google. But how can one find the information to analyze scientific, technical, economic and social factors affecting the innovative technical, technological, or financial mechanisms of coal-mining in the eastern regions of Russia?

To solve such search problems users have to employ lots of key concept combinations, clarify them in the course of en-route search on the Web or specialized stores such as patent databases (DB). It is not obvious that for this purpose any reasonable method would be used without fail. Eventually, a large amount of search results would be at the disposal of a user (tens and hundreds of documents), with the found information being more or less relevant to queries. As is quite common, there would be no opportunity to go into details of all the result data. So, the following questions can arise:

• How can one simultaneously assess the relevance of documents found by different queries? Is the relevance of documents determined correctly?

• Is the data ranking in a certain search system correct from the perspective of a user? Do all the results available for direct assessment meet the user's expectations?

• Are all the results that meet the user's expectations available for direct assessment? Are all the required data (e.g. innovative solutions) found at all?

• How one can filter documents extrinsic to the searched subject?

• Is it possible to find any effective solutions relevant in other application fields, but would be successfully used as an innovation in this domain.

• Is it possible to give a visual assessment to lots of found innovative solutions together with linked objects?

There are no clear-cut ways of solving these problems within trivial solutions. Obviously, we need efficient

This work was supported by the Russian Foundation of Basic Research (contract No NK13 -07- 00342)

methods of creating and populating the computer-assisted collections of advanced technologies and ideas which would contain not only their descriptions, but selected, classified and associated data. These data can be used to analyze retrospective and prospects of specific innovations, to search current and likely trends. The project in question is an attempt to offer a number of such innovative approaches.

### III. OUR APPROACH AND OTHER STUDIES

Currently, the R&D management assurance is of great importance (see, for example, the US National Trends and International Linkages in [1]). In this context, the automated semantic processing of large arrays of scientific and technical information, for sure, is used to search for breakthrough technologies and other innovative concepts, in the same manner as it is done in the prior art solutions illumin8 [2], NetBase [3], Orbit [4], Kalypso [5], as well as in large data stores such as CORDIS [6]. The intensive application of these and other similar tools make our attention focused on IT-related issues [7], [8].

The brief overview of publications, which are instrumental in pinning down the goal of the project under discussion, is given below. Our current development solutions mainly deal with the problem of data filtration [9] in terms of a content-based approach. In this respect it is important to note some interesting and topical visions of users and developers of RDF [10], semantic web-services control [11], as well as development of the document vector space model which is fundamental to most information retrieval problems, including rating of documents, data filtering, classification and clusterization of documents [12].

Taxonomy of web searches is also of great interest to us [13]. It is worth mentioning the pattern-searching procedure like the one recently described in [14] Information Filtering by Multiple Examples (IFME) which allows users to identify their information needs as a set of relevant documents, not keywords. The use of additional relevance assessment sources can help in work with lots of short texts (for example, Twitter). The use of alternate algorithms focused on solving the top-N recommendation task [15] seems to be useful too.

Another important field of our work is associated with effective query formulation. It is subject to optimal use of the combination of information sources in order to create an extended search set [16], [17]. The next essential study to be mentioned is [18]. It offers models and infrastructure for complex searches.

Note that our paper significantly expands and details [19].

### IV. PROBLEM DEFINITION

Thus, a project goal can be summarized as: the exploration of new approaches to innovative solution search methods in the database of a data center and its population with Internet data mining results adapted to visual assessment of selected, classified and associated data. We see three key tasks to attain the goal:

• To develop the technology of building and qualification of search queries with the following filtering and ranking of search data.

• To set up methods of cluster analysis to text documents and multimedia objects in order to use them for tagging the links between search results.

• To create a store of innovative solutions for educational and scientific purposes.

### V. SOFTWARE ARCHITECTURE

When developing a general software architecture based on mechanisms of direct automated search of innovative solutions the authors determined view layers, those of services, business logic, data access as well as crosscutting concerns (the UML notations and artifacts were applied). In the behavioral model of a system (Fig. 1), in a particular session, we can distinguish two periods of user's activation: query formulation (first step) and visualization of the results including the options of the requested and innovative solutions and linked objects (final step). Interim steps are hidden, off-line run and implement the algorithm of interaction between the system components without active participation of a user.

The main functional components:

• Search module. It involves executing a search query in the Internet search systems and the custom directory of innovative solutions; basic search (query by attributes and full-texts), location, data retrieval and summarizing.

• Query qualification module. Selection and ranking of search results: filtration, subject control, qualification of search query.

• Classification module. Classification of search results: selection of a method, cluster analysis of text documents and multimedia objects, data qualification. As a result we obtain a subset of semantically linked data.

• Link identification module. Link start-up: qualitative classification assessment, selection of the best results, interpretation of results; generating the descriptions of solutions with innovative potential in a given subject segment or for a specified object (article, technology, product).

• Visualization module. It involves mapping of search results, procedures of data processing, classification results including semantic links between objects.

• Data warehouse (DW) management module involves storage and updating of data search and processing results, parameters, and intermediate data; registry of innovative scientific, technological and educational problems. DW is built on the basis of a vector space model, includes document database access libraries and a data indexer.

• Service module. It involves monitoring and analysis of user access to information resources.

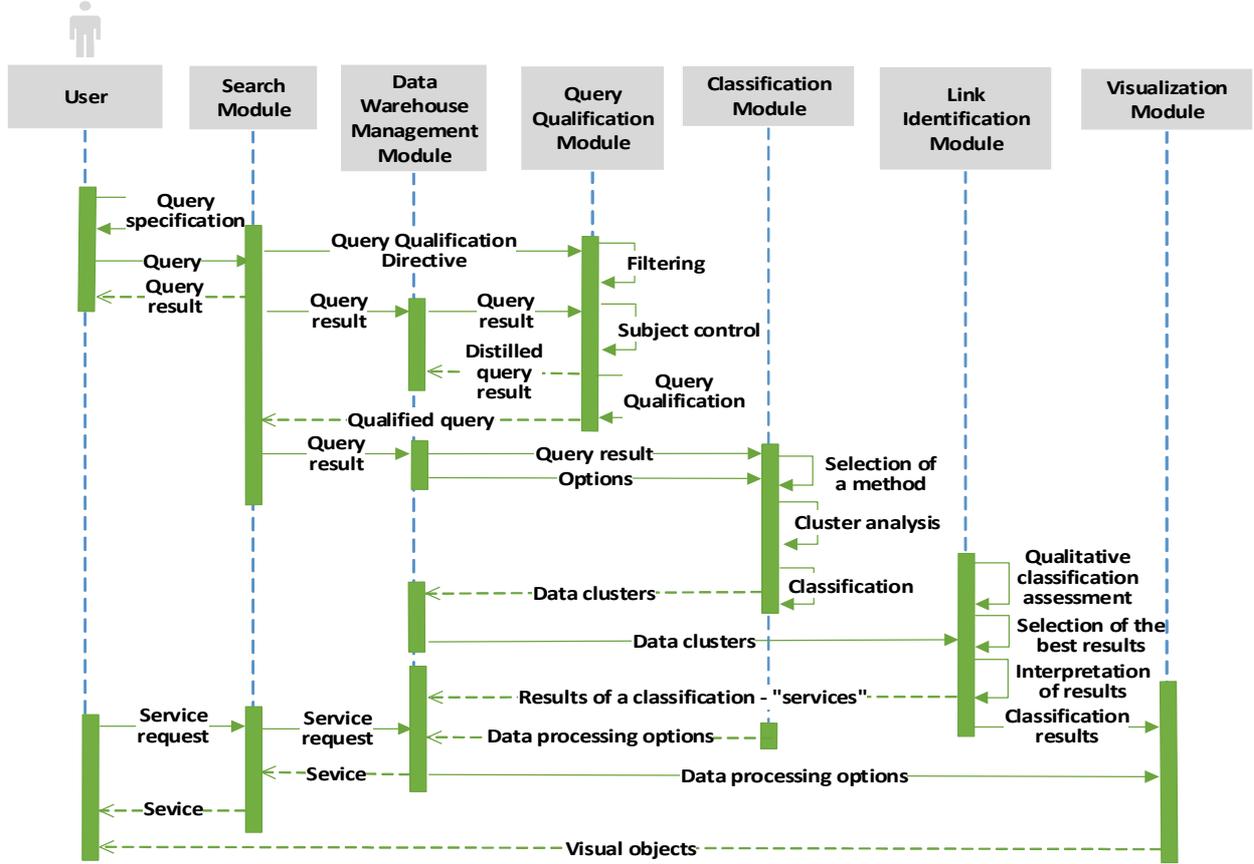

Fig 1. Behaviour of software components

It is particularly remarkable that the developed original object model is oriented to work with any text objects related to the subject of the processing: queries, search results, text documents. Over 30 entity classes specify a document processing environment, a set of documents, methods of calculating the package document similarity measures as well as search functions in document package, types of reports, a collection of document words, lemmatization, a document structure and its specific parts.

The detailed architectural solutions are described in [20].

## VI. GENERAL SEARCH ALGORITHM

One of the elements of the presented above architecture is a generalized heuristic algorithm for filtering and rating the search results, which is based on available search engines; the algorithm is supposed to provide a background for search modules and inquiry qualification, as well as for retrieval schedule and search procedures in general.

The algorithm under consideration uses search results of known search engines being in service; it is invariant to them; with various degree of automation; it uses the search engine rating results.

The algorithm instruments a multistep process of sequential filtrations of search results and the analysis of semantic similarity of the found object content to adaptively generated reference texts ($k$-patterns). Ranking quality of the filtered search results made as per algorithm was estimated by $DCG$ metric (see below). The ways of generating effective $k$-patterns were investigated as well.

Let us briefly run through the algorithm operation (Fig. 2). The description of a generalized request $Q_o$ includes the initial set of key concepts of the target document subject.

The generation of the set $Q$ of search queries $q \in Q$ and $|Q|=N$ is automated with an adaptive genetic algorithm searching for an effective total pertinence of the resulting document sampling under given evolutionary process depth constraints (see below).

The execution of queries $q_{ij}$ is accompanied with filtering search results $R_{qs}$ rated by a search engine and generating total results $R$. Filtering provides for the exclusion of some documents which subject area is formally pertinent but should not be the subject of the search for some reasons. It is done by hand or with a classifier which learning set is updated during the analysis of found texts.

The examples of documents being filtered are tutorials, student's papers, training programs, tests and notes, site promotion materials, company's sites, shopping sites, social networking sites; blogs; advertisements; virus-infected resources; nonexistent resources. The generation of k-patterns or reference texts is done simultaneously. They are used for calculating document similarity measures ($P_{ka}$ is a text combination based on the first positions of rated search results, $P_{kc}$ is the most pertinent result, $P_{kb}$ is the text constructed from authority dictionary entries and $P_{kd}$ is a text constructed from $Q_o$). Further the model of document vector space is used, i.e. each document $d$ (the search results from $R$ and k-patterns) is interpreted as vector $\bar{v}(d)=(w_{1,d}, w_{2,d}, w_{N,d})$, where $w_{t,d}$ is determined with a common metric $tf_{t,d} * idf_{t,d}$. A matrix $M_{N,x4}$ of document semantic similarity from set $R$ with common k-patterns is generated and the rating of documents from $R$ in accordance with their similarity $Similarity(d_1 d_1)$ to k-patterns is done.

The algorithm and some results of its use in patent searching are described in detail in [21]. To assess the algorithm quality, DCG metric [22] was used. For documents arranged by semantic similarity to $Similarity(d_1 d_1)$, the values were calculated for every k-pattern:

$$DCG = \sum_{p=1}^{n} 2^{gr(p)} - 1 / \log_2(2+p),$$

where $gr(p)$ is mean expert relevance assessment given to the document located on p position in the list of results, $gr \in [0,3] 10$ with 3 standing for "relevant", 0 – "irrelevant", 1 and 2 – "partially relevant" ("relevant (+)" or "relevant"); $1/\log_2(2+p)$ – document position discount (the documents at the head of the list are of greater importance).

Fig. 3 shows the ratio of DCG metrics of ideal ranging and various k-patterns. Good agreement of metric values is observed in various patterns. At the same time there are reserves for more exact labeling of documents by relevance groups gr. The algorithm under consideration labels 10-15% of documents as a group with value gr which is different from ideal ranging (see peaks of breakpoints). Then normalized values NDCG=DCG/Z were calculated for every k-pattern, with Z being equal to the greatest possible DCG value in case of ideal ranging according to the expert assessment. Indicator NDCG assumes values from 0 to 1. The ratio of NDCG values for generalized query and various k-patterns is presented in Fig. 4. It is evident that algorithm shows the best results under k-pattern $P_{ka}$ (combination of texts from the first positions of the ranked search results).

Note that the project provides for the usage of Internet search engine work results. The proposed search algorithms will be added to authoritative decisions – classical approaches to search result ranking (HITS, PageRank, BrowseRank, MatrixNet) which are based on the combination of document semantic pertinence and authority as well as user's behaviour and experience.

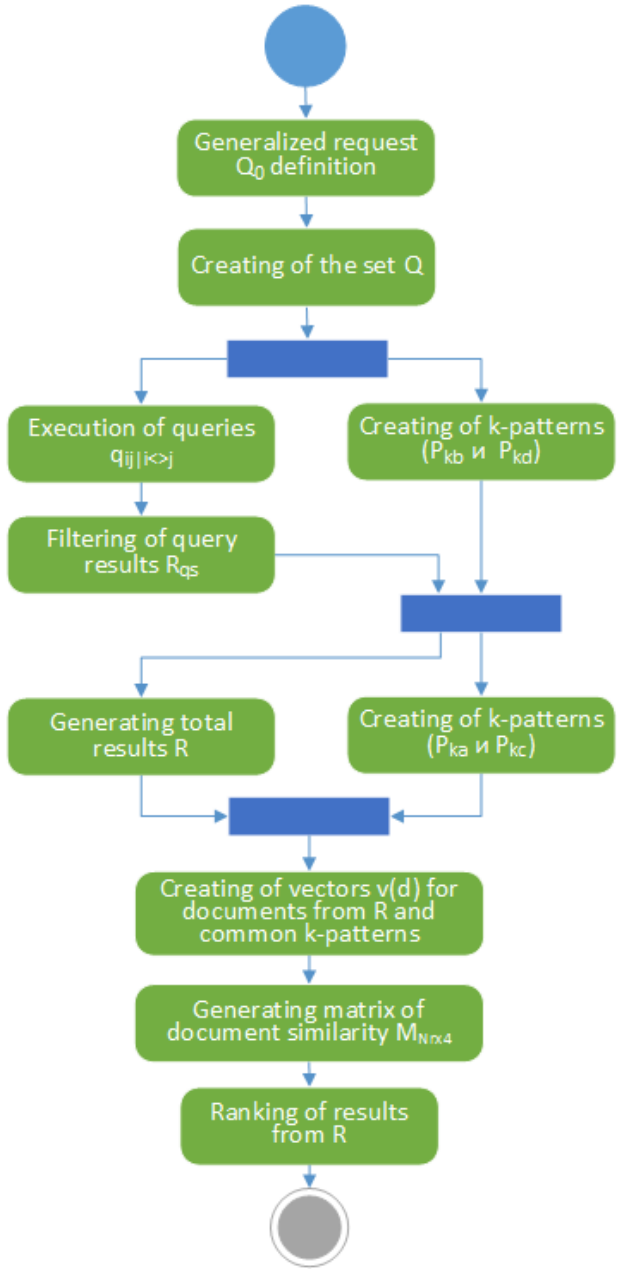

Fig 2. General pattern of a generalized heuristic algorithm for filtering and rating the search results

## VII. GENERATION OF SEARCH QUERIES

The project proposes and investigates the approach to search result generation based on a genetic algorithm (Fig. 5). The approach is used to specify a semantic kernel of a document desired set and generate sets of effective queries. The problem definition provides for the organization of an evolutionary process generating a stable and effective query population forming a relative search image of a document. A target set of search results is to be formed by such document addresses which are (a) in the first positions of a ranked list constructed by a search engine; (b) present in the result lists of multiple queries; (c) semantically similar to reference texts generated during evolutionary queries; (d) adequate to the environment given to a crawler by a user profile.

The original population from $N$ search queries may be a set of $Q=\{q_i\}$, $|Q|=N$, $N<|Q_0|/2$, $q_i=(k_1, k_2, ..., k_m)$, where $(k_1, k_2, ..., k_m)$ is a random combination of key concepts of a search image $Q_0$. The value of an objective function must determine the query quality (population individual fitness). For each $i$-th query result the value may be calculated as $w_i(f,p,s,a)$, where $f$ is determined by a result position in a ranked result list made by a search engine; $p$ is determined by entering the result in the result lists of most queries; $s$ is determined by a semantic similarity to $k$-patterns formed adaptively during the algorithm execution; $a$ is determined by a user profile as an environment factor (values $f,p,s,a$ are normalized for the range from 0 to 1). The value of a target function for each query is calculated as an averaged weight of query results, where $w_i$ is a weight of each result calculated after executing all queries; $P$ is a number of document addresses seen as the query result. The value of a objective function is interpreted as the capability of a search query to generate the results to be in the next population generation.

To choose parent couples the method of genotype outbreeding is proposed. It can provide for the most complete participation of all current queries in generating the next query population (the first parent individual is chosen randomly and the second individual is the "farthest" from the first one, the distance can be calculated as $\hat{w}=\bar{w}_1-\bar{w}_1$ ).

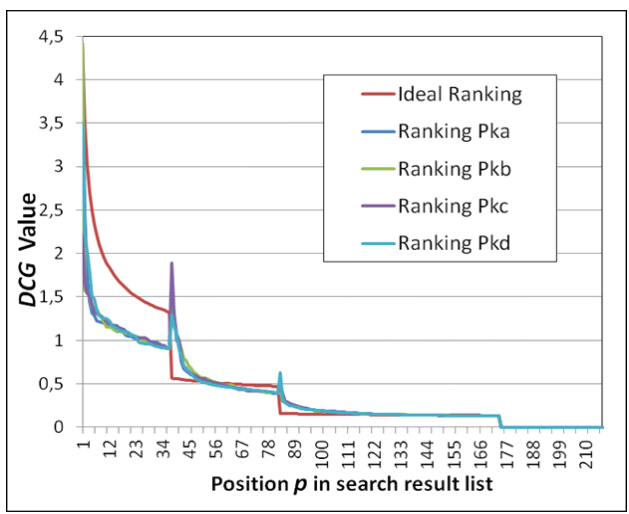

Fig 3. *DCG* values of ideal ranging and various *k*-patterns.

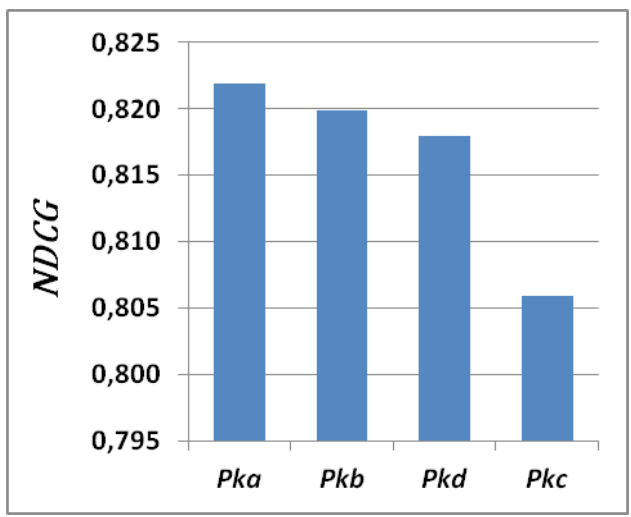

Fig 4. *NDCG* values of various *k*-patterns.

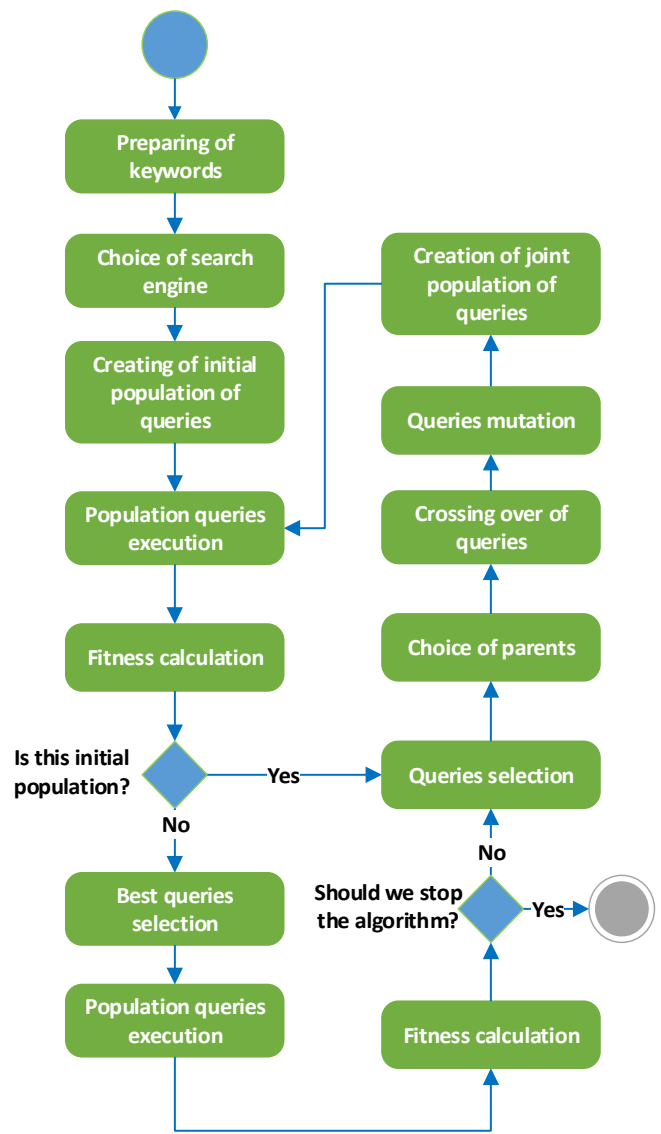

Fig 5. Genetic algorithm for search result generation.

The evolutionary operator of crossover is done with discrete recombination which corresponds to the exchange of key words (genes) between queries. The peculiarity of the proposed implementation is that the key word of a parent query is not substituted for the other parent query key word but its synonym. It allows generating considerably more child queries, with properties (semantics) of parent queries being preserved.

The essence of the most adequate mutation operation of the approach under study is the probabilistic change of a key query word (gene) chosen randomly. The essence of mutation of the approach under study is the change of a key query word (gene) chosen randomly. Since the number of key words in a query $q_i=(k_1, k_2, ..., k_m)$ is fixed, it is not possible to use such mutation operators as a new gene addition, new gene insertion, gene deletion. Besides, there is no sense in gene place exchange in the context of executing search queries.

To generate a new population an elite selection denying the loss of best solutions is used. An intermediate population is generated. It includes both the parents and their children. $N$ with the best values of a objective function $\hat{w}$ is chosen from all the population members. they will go in the next population. Generally, the condition of terminating the algorithm is considered to be population stability. For example, when a mean-square deviation of fitness function $\bar{W}$ reaches some threshold specified by an algorithm parameter. The genetic algorithm is described in detail in [23].

Some results of preliminary experimental studies of algorithm are briefly described below. The developers used original software support, search engine Bing and the following initial values of key parameters: $N$=15, $m$=3, $P$=10|50; number of search results returned after ranking all the results - 50. Weights of arguments $f$, $p$ and $s$ in search ranking were taken as equal to each other. The science of calculation of fitness function for groups of results is average value; the algorithm exit strategy is given number of passes. Terms from the document corpus (students' papers) were used in the origin collection .

Fig. 6 shows the plots of $\bar{W}$ against population number, with $P$=10 and $|Q_0|$=50 . Local maxima $\bar{W}$ and points of relative stabilization $\bar{W}$ (the 6-7th population) can be observed.

Fig. 7 shows plots of $\bar{W}$ against number of keywords of every generated query $m$ (shown as numbers beside plots). It is seen that the increase of $m$ leads, in the large, to the population quality improvement.

Fig. 8 shows the influence of fitness function arguments on its value. The greatest influence belongs to $f$, the lowest one – to $p$.

## VIII. Data Warehouse

The possibilities of Data Warehouse (DW) generation with realizing a document vector space model [24] to use it as a base of a data-centre information support are researched in the project. A software platform Document Text Analyzer (DTA) for semantic document analysis (their metric similarity computation) is developed within DW.

The prototype of the DW was tested successfully when associated technologies of the integral electronic document quality assessment and document pertinence in different contexts analysis were employed [25]. In particular, the debugging of software shell and interface of the TSTU specialized electronic teaching pack database, data centre warehouse components, was done (Fig. 9). The database is used to test and apply the project research results. The pioneering technology of the students' work uniqueness assessment (course and design-graphic papers, semester tasks, reports, essays, tests) is put to use.

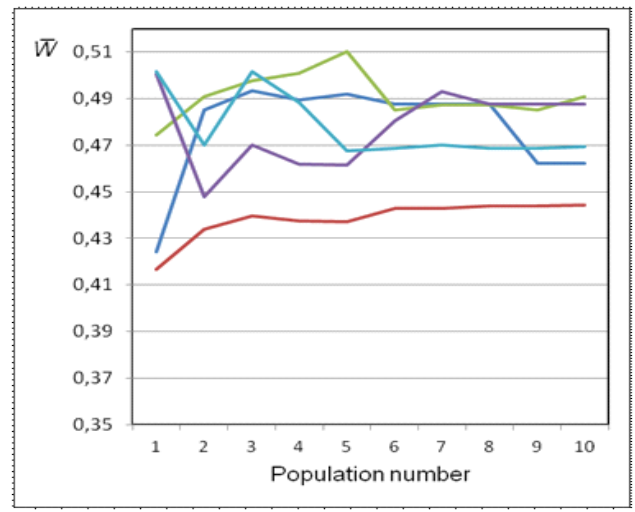

Fig 6. The plots of fitness function $\bar{W}$ against population number, $|Q_0|=50, m=3$ .

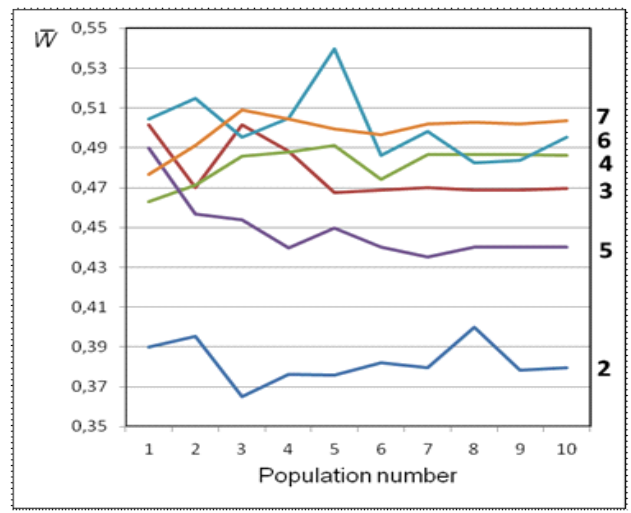

Fig 7. The plots of fitness function $\bar{W}$ against number of keywords of generated queries, $|Q_0|=50, m=2...7$

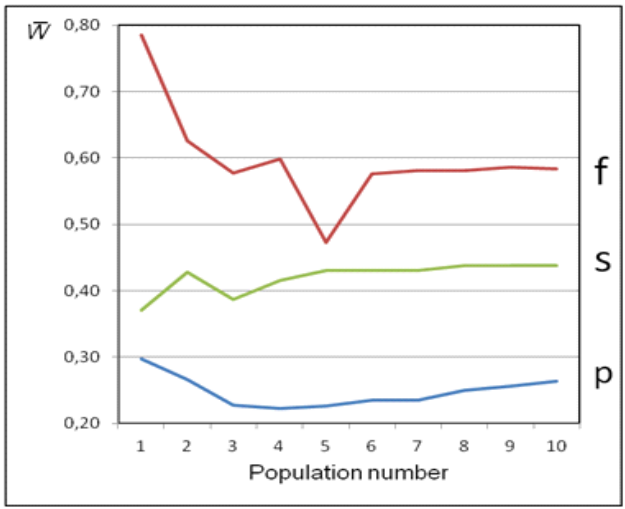

Fig 8. The influence of fitness function arguments on $\bar{W}$ value.

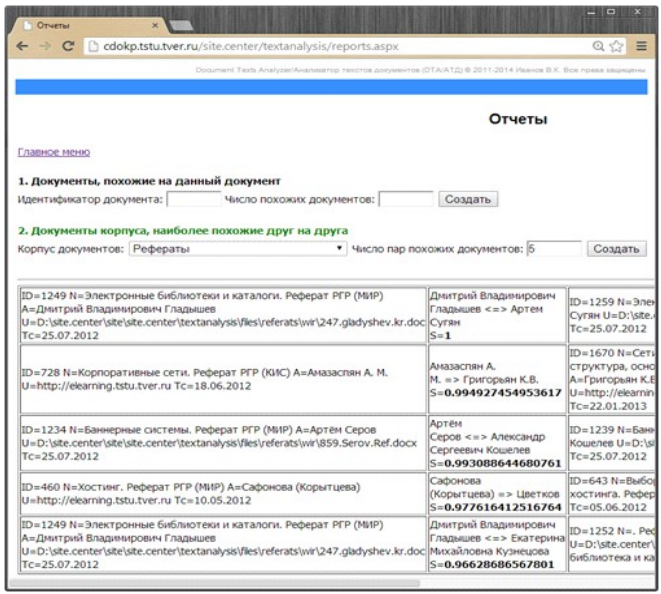

Fig 9. The appearance of one of the reports of experimental software platform DTA

The methods of semantic text comparison are used here. They are the computation of key concept weights and the construction of document vectors and not the known approaches (e.g. shingling) based on detecting direct adoptions in the text.

The research of some approaches to data centre different information systematization should be noted. As a result, a multistep algorithm of alternative search in an information catalogue with a target step number to be a base of a desired solution selection is developed [26].

IX. APPLICATION AREAS

The list of application areas of the approaches under discussion in the paper, the research results and technologies is given below:

• A competitive analysis and competitive intelligence. A survey of commercial, scientific and technical, social information sources in a target field. A search of business valuable information. A client information acquisition (in CRM systems). A characterization of new fields and directions in business planning. A search of sector innovation decision descriptions.

• Educational technologies. An analysis of students' paper works (graduation, course papers), theses. A selection and expert examination of teaching materials (books, articles, papers, essays, surveys, etc., including web-resources). Scientometrical analytical services.

• The work of competition committees and sponsoring agencies. An expert examination in venture and other investment funds, the work of councils and groups of experts. An analysis of applications, information cards, competition documentation, expert examination rules and conditions. Normalizing and metrological control of technical documentation. An analysis of project design documentation, standards, norms, rules, regulations, manuals.

• Patenting, novelty expert examination. Materials selection for patent investigations. A documentation analysis of intellectual property objects, license contracts. Technological development forecasting.

• A content analysis of document texts in sociological surveys.

• Staff recruitment at enterprises and in organisations. An analysis of applicants' resumes vacancy descriptions.

• Rubrication of personal digital documents. PC text document (files) classification and grouping.

It should be noted that the project made some patent research which aim was to find analogs of the system designed and establish its novelty. At the moment of the research result report preparation any data of direct project analogs or its components realized are not discovered. The search of the Federal Institute of Industrial Property's document database did not show any matches of the project results with technologies recorded in official publications of the titles of protection.

X. CONCLUSION

One of the R&D management reference models include a competitive analysis and technological development forecasting based on scientometrical analytical services and semantical systems of business valuable information search. A relatively new world trend is evident: an effective use of global knowledge dataflow. With all the differences the major search pattern is selecting materials on demand, highlighting key concepts in the desired area and grouping materials respectively, filtering and semantic result processing, generating analytical reports. In this sense, the project tasks the results of which were discussed in the article are timely and urgent, and on the appropriate level of the problem interpretation.

REFERENCES

[1] National Science Board. 2010. Science and Engineering Indicators 2010. Chapter 4. Research and Development: National Trends and International Linkages. Arlington, VA: National Science Foundation (NSB 10-01), 66 p.
[2] Illumin8. A powerful research tool for innovation & product development. URL: http://www.illumin8.com.
[3] NetBase Social Media Management System (SMMS). URL: http://www.netbase.com.
[4] Questel Intellectual Property Portal. URL: http://www.orbit.com.
[5] R&D Management Framework. URL: http://kalypso.com/rd.
[6] Community Research and Development Information Service (CORDIS). URL: http://cordis.europa.eu.
[7] Nambisan S. (ed.). Information Technologies and Product Development, Annals of Information Systems, Vol. 5, 218 p. DOI 10.1007/978-1-4419-1081-3_1.
[8] Song L.Z., Song M. The Role of Information Technologies in Enhancing R&D–Marketing Integration: An Empirical Investigation. Journal of Product Innovation Management, Vol. 27, Issue 3, pp. 382–401, May 2010. DOI: 10.1111/j.1540-5885.2010.00723.x.
[9] Hanani U., Shapira B., Shoval P. Information Filtering: Overview of Issues. Research and Systems, User Modeling and User-Adapted Interaction II: pp. 203-259. 2001.
[10] Fensel D., Patel-Schneider P.F., Layering the Semantic Web: Problems and Directions. ISWC'02 Proceedings of the First International Semantic Web Conference on The Semantic Web, pp. 16-29, 2002, DOI: 10.1007/3-540-48005-6_4.
[11] Dong H., Hussain F.Kh., Chang E. Semantic Web Service matchmakers: state of the art and challenges. Concurrency and Computation: Practice and Experience, Vol. 25, Issue 7, pp. 961–988, May 2013, DOI: 10.1002/cpe.2886
[12] Manning C. D., Raghavan P., Schütze H.. Introduction to information retrieval. Cambridge University Press, Cambridge, England, 2008, 482 p.
[13] Broder A. A taxonomy of web search. ACM SIGIR Forum Vol. 36, Issue 2, Fall 2002, pp. 3 – 10, DOI: 10.1145/792550.792552.
[14] Zhu M., Xu C., Wu Y.-F.B. IFME: information filtering by multiple examples with under-sampling in a digital library environment. JCDL'13 Proceedings of the 13th ACM/IEEE-CS joint conference on Digital libraries, pp. 107-110, DOI: 10.1145/2467696.2467736.
[15] Cremonesi P., Koren Y., Turrin R. Performance of recommender algorithms on top-n recommendation tasks. RecSys'10 Proceedings of the fourth ACM conference on Recommender systems, pp. 39-46, DOI: 10.1145/1864708.1864721.
[16] Wu J., Ilyas I., Weddell G. A Study of Ontology-based Query Expansion, Cheriton School of Computer Science, University of Waterloo, Technical Report CS-2011-04, February 09, 2011, p. 38
[17] Bendersky M., Metzler D., Croft W. B. Effective query formulation with multiple information sources. WSDM'12 Proceedings of the fifth ACM international conference on Web search and data mining, pp. 443-452, DOI: 10.1145/2124295.2124349.
[18] Ageev M., Guo O., Lagun D., Agichtein E. Find it if you can: a game for modeling different types of web search success using interaction data. SIGIR'11 Proceedings of the 34th international ACM SIGIR conference on Research and development in Information Retrieval, pp. 345-354, DOI: 10.1145/2009916.2009965.
[19] Ivanov, V.K., Palyukh, B.V., Sotnikov, A.N. Intelligent subject search support in science and education. Innovative Information Technologies : Materials of the III International scientific-practical conference. Part 2. Innovative Information Technologies in Science / Ed. S.U. Uvaysov. - pp. 34-40. - M., 2014.
[20] Ivanov V.K., Palyukh B.V., Sotnikov A.N. Arkhitektura intellektual'noy sistemy informatsionnoy podderzhki innovatsiy v nauke i obrazovanii // Programmnyye produkty i sistemy. – Tver', 2013. – № 4. – pp. 197-202.
[21] Ivanov V.K., Vinogradova N.V. Evristicheskiy algoritm fil'tratsii i semanticheskogo ranzhirovaniya rezul'tatov poiska dokumentov // Vestnik Tverskogo gosudarstvennogo universiteta: Seriya "Prikladnaya matematika" / № 41. –Tver', 2013. – № 3. – pp. 97-107.
[22] Järvelin K., Kekäläinen J. Cumulated gain-based evaluation of IR techniques. ACM Transactions on Information Systems (TOIS), Vol. 20, issue 4, October 2002, 422-446, DOI: 10.1145/582415.582418.
[23] Ivanov V.K. Osnovnyye shagi geneticheskogo algoritma fil'tratsii rezul'tatov tematicheskogo poiska dokumentov: stat'ya // Innovatsii v nauke. – Novosibirsk, 2013. –№ 25. – P. 8-15.
[24] Salton G., Wong A., Yang C.S. A Vector Space Model for Automatic Indexing. Communications of the ACM. - 1975. - Vol. 18, nr. 11. - pp. 613–620.
[25] Ivanov V.K., Mironov V.I. Osobennosti analiza skhodstva dokumentov v razlichnykh kontekstakh zaimstvovaniya pri podgotovke tekstovykh materialov // Otsenka kachestva vysshego professional'nogo obrazovaniya s uchetom trebovaniy FGOS i professional'nykh standartov: materialy dokladov – Tver', 2013. – pp. 19-26.
[26] Paliukh, B., Egereva, I. Multistep algorithm of alternatives search in an information catalogue // 10th International Conference on Interactive Systems: Problems of Human-Computer Interaction. – Collection of scientific papers. – Ulyanovsk : USTU, 2013. – pp.. 129-132.